%% file: jpsi276-alicepreprint_031012.tex
\documentclass[ALICE,manyauthors]{cernphprep}
\usepackage{cite}
\newcommand{\jpsi}{\rm J/$\psi$}

\usepackage{rotating}

\begin{document}%
%
%
\begin{titlepage}
\PHnumber{2012-055}      
\PHdate{v3, 19 October 2012}              
%
%
\title{Inclusive 
J/$\mathbf{\psi}$ production
in pp collisions at $\mathbf{\sqrt{s}} = 2.76$ TeV}
\ShortTitle{Inclusive 
J/$\psi$ production
in pp collisions at $\sqrt{s} = 2.76$ TeV}   
%
\Collaboration{The ALICE Collaboration}%
\ShortAuthor{The ALICE Collaboration}      
\begin{abstract}
The ALICE Collaboration has measured inclusive J/$\psi$ production in pp collisions at a center of mass energy $\sqrt{s}=2.76$ TeV at the LHC. 
The results presented in this Letter refer to the rapidity ranges 
$|y|<0.9$ and $2.5<y<4$ and have been obtained by measuring the electron and muon pair decay channels, respectively. The integrated luminosities for the 
two channels are  $L^{\rm e}_{\rm int}=1.1$~nb$^{-1}$ and 
$L^{\mu}_{\rm int}=19.9$~nb$^{-1}$, and the corresponding signal statistics are
$N_{\rm J/\psi}^{{\rm e}^+{\rm e}^-}=59\pm 14$ and 
$N_{\rm J/\psi}^{\mu^+\mu^-}=1364 \pm 53$.
We present ${\rm d}\sigma_{\rm J/\psi}/{\rm d} y$ for the two rapidity regions under study and, for the forward-$y$ range, ${\rm d}^2\sigma_{\rm J/\psi}/{\rm d}y{\rm d}p_{\rm t}$ in the transverse momentum domain $0<p_{\rm t}<8$ GeV/$c$. The results are 
compared with previously published results at $\sqrt{s}=7$ TeV and with theoretical calculations.
\end{abstract}
\end{titlepage}
%
\input{jpsi276_031012.tex}               

\input{references.tex}
%
\newenvironment{acknowledgement}{\relax}{\relax}
\begin{acknowledgement}
\section*{Acknowledgements}
\input{acknowledgements_Jan2012.tex}    
\end{acknowledgement}
%
%
%
\appendix
\section{The ALICE Collaboration}
\label{app:collab}
\input{authorlist_17Feb2012.tex}  
\end{document}

%% file: jpsi276_031012.tex
\section{Introduction}

Almost forty years after the discovery of charmonium, its production in hadronic collisions still remains not completely understood, and charmonium production data represent a complex and severe test for QCD-inspired models~\cite{QWG11}. 

Recently, first results from the Large Hadron Collider (LHC) on J/$\psi$ production in pp collisions at $\sqrt{s}=7$~TeV became available~\cite{Kha10,Aai11,Aad11,Aam11,Cha11}, significantly extending the energy reach beyond that of the Tevatron and RHIC hadron colliders~\cite{Aco05,Abb99,Ada07}. A reasonable description of the transverse momentum spectra has been obtained by theoretical models~\cite{But11,Lan09,Ma11,Fra08}, and first results on J/$\psi$ polarization, a crucial testing ground for theory~\cite{Abu07,Fac09,Hab08}, are also available~\cite{Abe11} at LHC energy.

At the beginning of 2011, the LHC delivered pp collisions at $\sqrt{s}=2.76$ TeV. The main goal of this short run was to provide a
reference for the Pb--Pb data which were taken at the same $\sqrt{s}$ per nucleon-nucleon collision. On the other hand, these data offer the possibility of studying J/$\psi$ production at an intermediate energy between Tevatron and the present LHC top energy, and represent therefore an interesting test for models.

In this Letter, we present results on inclusive J/$\psi$ production at $\sqrt{s}=2.76$ TeV as obtained by the ALICE experiment~\cite{Aam08}. J/$\psi$ particles were measured, down to zero transverse momentum, via their decay into e$^+$e$^-$ at mid-rapidity ($|y|<0.9$) and into  
$\mu^+\mu^-$ at forward rapidity ($2.5<y<4$).
Results from ALICE on J/$\psi$ production at $\sqrt{s}=7$ TeV were recently published~\cite{Aam11,Abe11}. Since the experimental apparatus and the data analysis procedure are basically the same for the two data samples, they will be concisely described, referring where necessary to our previous publications. Results will then be shown for ${\rm d}\sigma_{\rm J/\psi}/{\rm d} y$ at central and at forward rapidity. For the region $2.5<y<4$ the differential cross section  ${\rm d}^2\sigma_{\rm J/\psi}/{\rm d} y{\rm d} p_{\rm t}$ will also be given, for the transverse momentum range $0<p_{\rm t}<8$~GeV/$c$. A comparison with the previous results at $\sqrt{s}=7$ TeV will be carried out and next-to-leading order Non-Relativistic QCD (NLO NRQCD)  theoretical calculations will be compared to the experimental data.

\section{Experimental apparatus and data analysis}

The main elements of the ALICE experiment at the CERN LHC are a central 
rapidity barrel (covering the pseudorapidity range $|\eta|<0.9$) for the detection of hadrons, electrons and photons and for the measurement of jets, and a forward muon spectrometer  ($-4<\eta<-2.5)$.
The experimental set-up is described in detail in~\cite{Aam08}.
For the analysis described in this Letter, the relevant detector systems for tracking and electron identification in the central barrel are the Inner Tracking System (ITS)~\cite{Aam10}, based on six layers of silicon detectors, and the Time Projection Chamber (TPC)~\cite{Alm10}. The ITS covers the $|\eta|<0.9$ range and, together with two small forward scintillator detectors (VZERO, covering $2.8<\eta<5.1$ and $-3.7<\eta<-1.7$), is used to define the Minimum-Bias (MB) interaction trigger. In particular, the MB condition requires a logical OR between at least one fired read-out chip in one of the two inner layers of the ITS (Silicon Pixel Detector), and a signal in at least one of the VZERO detectors. 
Muons are tracked and identified in the muon spectrometer~\cite{Aam11}, which consists of a front absorber to remove hadrons, a 3 T$\cdot$m dipole magnet and a tracking system. It also includes a triggering system with a programmable $p_{\rm t}$ threshold. With this trigger, the collected data sample was enriched with events where, in addition to the MB condition, at least one muon was detected 
in the spectrometer acceptance. The threshold for the muon trigger was set to its minimum value, $p_{\rm t}=0.5$ GeV/$c$. With this choice the acceptance for J/$\psi\rightarrow\mu^+\mu^-$ detection extends down to $p_{\rm t}=0$. Further details on the detectors relevant for this analysis and on the trigger definitions can be found in Ref.~\cite{Aam11}.

The dielectron analysis is based on a sample of 65.4$\cdot$10$^6$ MB triggers, corresponding to an integrated luminosity $L^{\rm e}_{\rm int}=1.1$ nb$^{-1}$. Out of the total sample, 47.4$\cdot$10$^6$  events have a reconstructed vertex which lies within $\pm$10 cm, along the beam direction, from the nominal interaction point and are retained for the following analysis steps.
The analysis strategy is briefly described below. It is the same as applied in case of the analysis at $\sqrt{s}=7$\,TeV, small differences are explained in the text. For details we refer to~\cite{Aam11}.

Reconstructed tracks are required to have a hit in one of the two innermost or in the fifth ITS layer (layers three and four were excluded from the reconstruction). 
This choice makes the track cuts somewhat less stringent as compared to the analysis of the $\sqrt{s}=7$\,TeV data where a hit was required in one of the two innermost layers. As a result, the signal increases by $\sim$12\%, whereas the significance for the two cuts is comparable within the uncertainties. The choice to use the looser cut was motivated by the fact that it provides a central cross section value of the systematic variations using different cuts.
The number of TPC clusters for each track must be larger than 70 (out of a maximum of 159), with the $\chi^2$ per space point of the momentum fit lower than 4. The kinematic cuts $p_{\rm t}>1$~GeV/$c$ and $|\eta|<0.9$ are applied to each track. The electron identification is based on the correlation between the specific energy loss ${\rm d}E/{\rm d}x$ and the momentum measured in the TPC, requiring a $\pm 3\sigma$ inclusion cut around the electron line corresponding to the Bethe-Bloch expectation and an exclusion cut of $\pm 3.5\sigma$ ($\pm 3\sigma$) for pions (protons). Finally, a rapidity cut $|y|<0.9$ is applied to to J/$\psi$ candidates to remove pairs at the edge of the acceptance.

\begin{figure}[htbp]
\centering{\includegraphics[width=.5\textwidth]{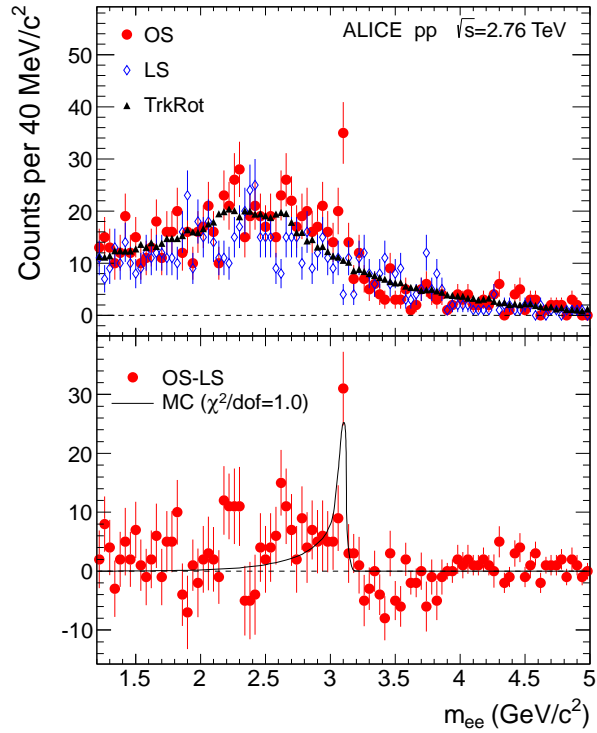}}
\caption{Top panel: invariant mass distributions for opposite-sign (OS) and 
like-sign (LS) electron pairs ($|y|<0.9$, all $p_{\rm t}$). The background estimate from the TrkRot method (see text for details) is also shown.
Bottom panel: the difference of the OS 
and LS distributions with the normalized MC signal shape superimposed.}
\label{fig:1}
\end{figure}

The signal extraction is based on the like-sign (LS) subtracted invariant mass spectrum of e$^+$e$^-$ pairs. The LS spectrum is obtained as the sum of positive-positive and negative-negative spectra. The scale factor on the LS background, applied in~\cite{Aam11} to account for various non-combinatorial effects, was found to be negligible in this analysis. Figure~\ref{fig:1} (top panel) shows the opposite sign (OS) dielectron mass spectrum together with the LS spectrum. After subtraction, the number of J/$\psi$ is estimated by bin counting in the invariant mass range 
$2.92 < m_{e^{+} e^{-}} < 3.20$\,GeV/$c^2$, resulting in $59 \pm 14$ (stat.) counts with a significance of $5.4 \pm 0.6$. 
The signal fraction in the mass range defined above is estimated from a Monte Carlo (MC) simulation, and included in the acceptance. 
In Fig.~\ref{fig:1}  (bottom panel) the LS-subtracted spectrum is overlayed with the MC signal shape, normalized to the data points in the invariant mass range $2.5 < m_{e^+e^-} <3.5 $\,GeV/$c^2$. In addition to the LS method, the background estimated using a track rotation (TrkRot) technique\footnote{In the TrkRot method one track of the OS pair is rotated around the $z$-axis. The procedure is repeated several times randomly varying the rotation angle. In this way,one removes the correlation between the two electrons of the pair. The TrkRot invariant mass spectrum is scaled to match the integral of the OS spectrum in the region $3.2<m_{{\rm e}^+{\rm e}^-}<5$ GeV/$c^2$.} is also shown in Fig.~\ref{fig:1}. The differences between the number of J/$\psi$ obtained with the TrkRot and LS methods is used in the estimate of the systematic uncertainty on the signal extraction.

The dimuon analysis is based on 8.8$\cdot$10$^6$ muon-triggered events, corresponding to an integrated luminosity $L^{\mu}_{\rm int}=19.9$~nb$^{-1}$. Out of this sample, 1.0$\cdot$10$^5$ events contain a reconstructed OS muon pair. It is required that each event contains at least one reconstructed vertex. Events are retained for the analysis if both candidate muon tracks exit the front hadron absorber ($z=-503$ cm) at a radial coordinate $17.6<R_{\rm abs}<89.5$ cm, a cut roughly corresponding to the angular acceptance of the muon spectrometer. It is also required that at least one of the two muons satisfies the muon trigger condition. 
Finally, the cut $2.5<y<4$ is applied to the pairs in order to reject dimuons at the edge of the spectrometer acceptance.

\begin{figure}[htbp]
\centering
\resizebox{0.6\textwidth}{!}
{\includegraphics*[bb=0 0 565 544]{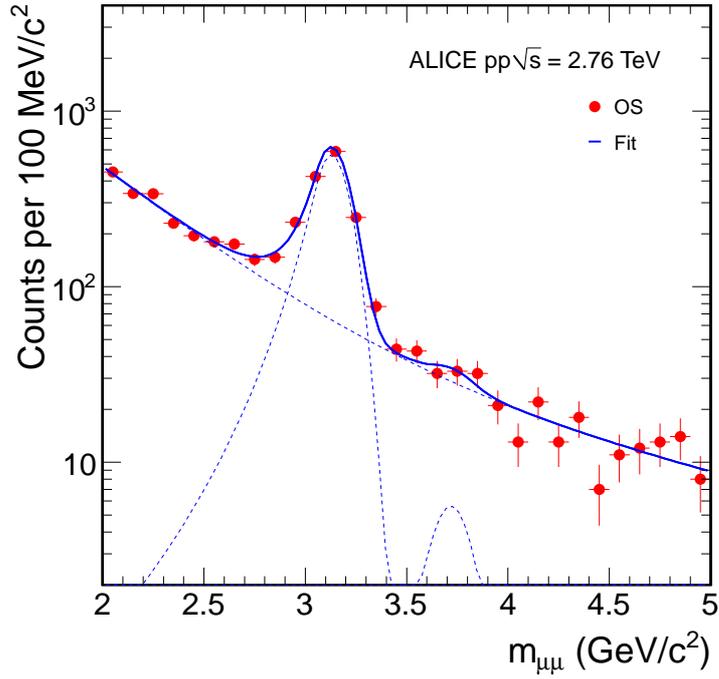}}
\caption{Invariant mass distribution for opposite-sign muon pairs ($2.5<y<4$, all $p_{\rm t}$), 
in the mass region $2<m_{\mu\mu}<5$ GeV/$c^2$, with the result of
the fit (see text for details). The fitted J/$\psi$ and $\psi(2S)$ contributions, as well as the 
background, are also shown.}
\label{fig:2}
\end{figure}

The signal is extracted by a fit to the invariant mass spectrum over the range $2<m_{\mu\mu}<5$ GeV/$c^2$. The signal is parameterized with a Crystal Ball (CB) function~\cite{Gai82} with a background described by the sum of two exponentials. The position ($m_{\rm J/\psi}$) of the peak of the CB function, as well as its width ($w_{\rm J/\psi}$), are kept as free parameters in the fit. The obtained values are $m_{\rm J/\psi}=3.129\pm 0.004$ GeV/$c^2$ (a value larger by  $\sim$1\% than the world average~\cite{Nak10}) and $w_{\rm J/\psi}=0.083\pm 0.004$ GeV/$c^2$.
The J/$\psi$ width is only slightly larger (by $\sim 0.006$ GeV/$c^2$) than that  obtained in the MC, which includes the effect of the misalignment of the muon tracking system. The tails of the CB function are fixed to their MC value, since with the available statistics and signal to background ratio they cannot be reliably extracted from the fitting procedure. Finally, the contribution of the $\psi(2S)$ signal is included in the fit, although its influence on the number of detected J/$\psi$ is negligible.
In Fig.~\ref{fig:2} the dimuon invariant mass spectrum is presented, together with the result of the fit ($\chi^2/ndf=1.3$). By integrating the CB function, one gets a total number of J/$\psi$ 
$N_{\rm J/\psi}^{\mu^+\mu^-}=1364\pm 53$~(stat.).

\begin{figure}[htbp]
\centering
\resizebox{1.0\textwidth}{!}
{\includegraphics*[bb=0 0 568 327]{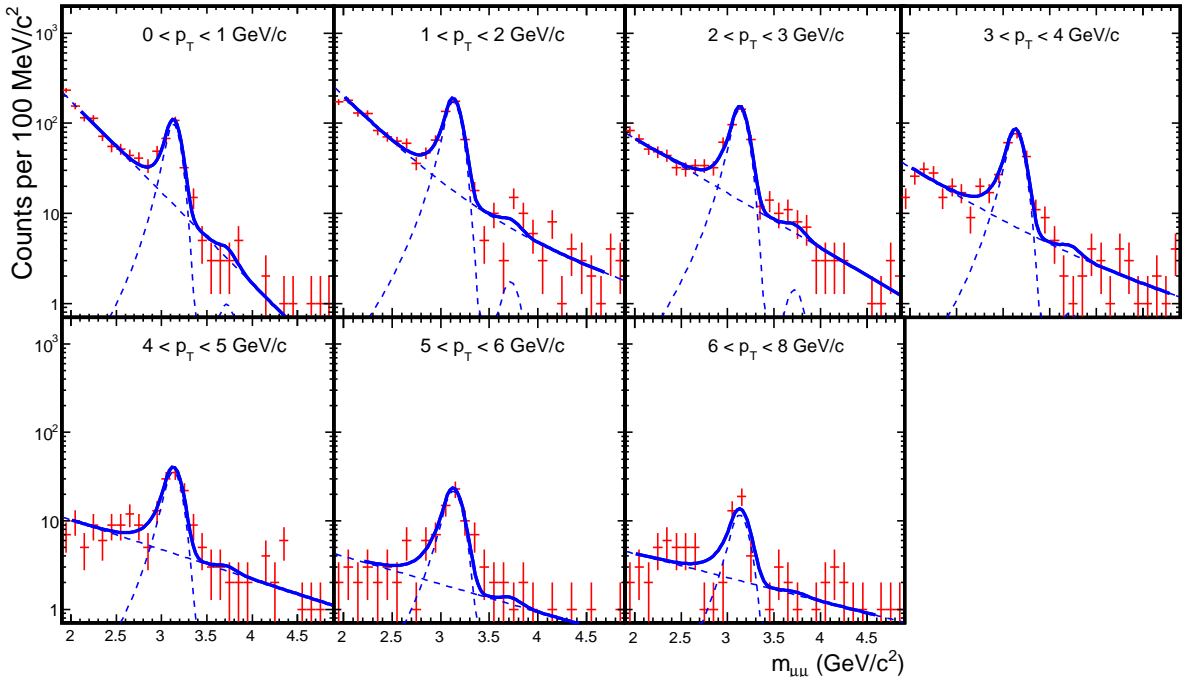}}
\caption{Invariant mass spectra for OS muon pairs ($2.5<y<4$), in bins of $p_{\rm t}$. The results of the fits are also shown.}
\label{fig:3}
\end{figure}

The J/$\psi$ statistics in the dimuon channel permit a differential study of the production cross sections using six $y$ or seven $p_{\rm t}$ intervals. The fitting technique is the same as for the integrated invariant mass spectrum, except for the value of the CB width which was fixed for each bin $i$ to the value $w_{\rm J/\psi}^i = w_{\rm J/\psi}\cdot(w_{\rm J/\psi}^{i,MC}/w_{\rm J/\psi}^{MC})$, i.e., by scaling the measured width for the integrated spectrum with the MC ratio between the widths for the bin $i$ and for the integrated spectrum. The sum of the signal events for both $p_{\rm t}$ and $y$ bins agrees well (within 0.3\% and 1.2\% respectively) with the result of the fit to the integrated mass spectrum.
In Fig.~\ref{fig:3} the invariant mass spectra corresponding to the various  $p_{\rm t}$ bins are shown, together with the results of the fits. The 
J/$\psi$ signal is well visible also in the spectra with lower statistics and the quality of the fits is similar to the one obtained for the integrated mass spectrum. 

For both the dielectron and dimuon analyses the number of signal events is corrected by the product of acceptance times efficiency ($A\times\epsilon$). 
The $A\times\epsilon$ values are obtained using MC simulations which include a description of the status of the detector as a function of time. Details on the procedure are given in Ref.~\cite{Aam11}.
For this analysis, the MC input distributions in transverse momentum and rapidity are obtained by interpolating between the LHC results for $\sqrt{s}=7$~TeV and lower energy collider measurements~\cite{Bos11}. It was verified a posteriori that the interpolated input spectra
are in good agreement with those obtained from this analysis. The results are $A\times\epsilon=0.120$ for the dielectron analysis and $A\times\epsilon=0.346$ for the dimuon analysis. These values refer to J/$\psi$ production for $p_{\rm t}>0$ in the analyzed rapidity ranges, $|y|<0.9$ and $2.5<y<4$, respectively. 

The inclusive J/$\psi$ production cross section for the leptonic channel $\ell^+\ell^-$ is calculated as:

\begin{equation}
\sigma_{\rm J/\psi} =  \frac{N^{\rm {cor},\ell^+\ell^-}_{\rm J/\psi}}
{BR(\rm J/\psi\rightarrow \ell^+\ell^-)}\times 
\frac{\sigma_{{\rm MB}}}{N_{\rm{MB}}}\times R^{ \ell^+\ell^-}
\label{eq:1} 
\end{equation} 

where $N^{\rm {cor},\ell^+\ell^-}_{\rm J/\psi}=
N^{\ell^+\ell^-}_{\rm J/\psi}/(A\times \epsilon)^{\ell^+\ell^-}$
is the number of signal events corrected for acceptance times efficiency,  
$BR(\rm J/\psi\rightarrow \ell^+\ell^-)=(5.94\pm 0.06)$\%~\cite{Nak10} is the leptonic branching ratio for the J/$\psi$ decay, $N_{\rm{MB}}$ is the number of MB-triggered events and $\sigma_{{\rm MB}}=55.4\pm 1.0$(total) mb is the absolute cross section for the occurrence of the MB condition~\cite{Gag11}, derived from the result of a van der Meer scan (see \cite{Aam11} for details). The 
$R^{\ell^+\ell^-}$ factor is 1 for the e$^+$e$^-$ analysis, whereas for the dimuon channel $R^{\mu^+\mu^-}=0.0326\pm 0.0002$ represents the inverse of the enhancement factor of the muon trigger with respect to the MB trigger~\cite{Aam11}.
An equivalent formula is used for the differential cross sections in $y$ and $p_{\rm t}$.   

The sources of systematic uncertainties are exactly the same as for the corresponding $\sqrt{s}=7$~TeV analysis and have been estimated in a similar way (see \cite{Aam11} for details). In Table~\ref{tab:1} we quote their values for the integrated cross sections in the dielectron and in the dimuon channel. The uncertainty on signal extraction for the electron analysis (14\%) is larger than the 8.5\% quoted at $\sqrt{s}=7$  TeV~\cite{Aam11}. This increase mainly comes from the difference in $N^{\rm {cor},{\rm e}^+{\rm e}^-}_{\rm J/\psi}$ obtained by requiring various conditions in the ITS: a hit in the first layer, in any of the first two layers (as was done for the $\sqrt{s}=7$ TeV analysis), or the less stringent condition adopted from the present analysis, described earlier in this Section. For the muon analysis, the uncertainty on signal extraction (4\%) is now smaller with respect to the 7.5\% quoted at $\sqrt{s}=7$ TeV~\cite{Aam11}. The present value was calculated as the average absolute deviation on the number of signal events obtained with alternative parameterizations of the signal and background shapes. At $\sqrt{s}=7$ TeV the more conservative, but also more prone to statistical effects, 
approach of using the larger deviation obtained in the various fits was adopted. Finally, the decrease of the systematic uncertainty on the trigger efficiency for the muon analysis (from 4\% at $\sqrt{s}=7$ TeV~\cite{Aam11} to 2\% at $\sqrt{s}=2.76$ TeV) is due to a different approach, the present one being based on the study of the variation of the J/$\psi$ triggering efficiency when the efficiency of the trigger detectors is changed by an amount slightly larger than the uncertainty on this last quantity.

The total systematic uncertainties, excluding those related to the unknown degree of polarization of the J/$\psi$, are 18.0\% and 8.1\% for the dielectron and the dimuon channel, respectively.
For the differential cross sections measured in the dimuon channel, the same sources of systematic uncertainties quoted in Table~\ref{tab:1} apply to each $y$ and $p_{\rm t}$ bin. For the uncertainties relative to the choice of the MC inputs, their values may in principle vary with either rapidity or transverse momentum. However, no clear trend as a function of these two variables is observed. So, the relative systematic uncertainty calculated for the integrated cross section is assigned to each point and considered as uncorrelated between the bins. The uncertainty on signal extraction is also considered as bin-to-bin uncorrelated. The limited signal statistics for most of the bins prevents a direct study of the systematic uncertainty, therefore the relative systematic uncertainty assigned to the integrated cross section was attributed to each point. 

\begin{table}
\caption{\label{tab:1} Systematic uncertainties (in percent) contributing to the measurement of the integrated \jpsi\ cross section. The uncertainties related to the J/$\psi$ polarization were calculated for both Collins-Soper and helicity reference frames.}
\centering
\begin{tabular}{c|c|c|c|c}
\hline
\hline
Channel & \multicolumn{2}{c} {e$^+$e$^-$} & \multicolumn{2}{|c} { $\mu^+\mu^-$} \\ \hline
Signal extraction & \multicolumn{2}{c} {14} & \multicolumn{2}{|c} {4} \\ \hline
Acceptance input & \multicolumn{2}{c} {1.5} & \multicolumn{2}{|c} {4} \\ \hline
Trigger efficiency & \multicolumn{2}{c} {$-$} & \multicolumn{2}{|c} {2} \\ \hline
Reconstruction efficiency & \multicolumn{2}{c} {11} & \multicolumn{2}{|c} {4} \\ \hline
R factor & \multicolumn{2}{c} {$-$} & \multicolumn{2}{|c} {3} \\ \hline
Luminosity & \multicolumn{2}{c} {1.9} & \multicolumn{2}{|c} {1.9}\\ \hline
B.R. & \multicolumn{4}{c} {1} \\ \hline
Polarization & $\lambda=-1$ & $\lambda=1$ & $\lambda=-1$ & $\lambda=1$ \\ \hline
CS & +19 & --13 & +32 & --16 \\ \hline
HE & +21 & --15 & +24 & --12 \\ \hline
\hline
\end{tabular}
\end{table}

\section{Results}

The analysis described in the previous Section gives the following results for the integrated inclusive J/$\psi$ cross sections in the two rapidity ranges investigated at $\sqrt{s}=2.76$~TeV:

$\sigma_{\rm J/\psi}(|y|<0.9)$~=~7.75~$\pm$~1.78(stat.)~$\pm$~1.39(syst.)~$+$~1.16($\lambda_{\rm{HE}}=1$)~$-$~1.63($\lambda_{\rm{HE}}=-1$)~$\mu$b and \\
$\sigma_{\rm J/\psi}(2.5<y<4)$~=~3.34~$\pm$~0.13(stat.)~$\pm$~0.27(syst.)~$+$~0.53($\lambda_{\rm{CS}}=1$)~$-$~1.07($\lambda_{\rm{CS}}=-1$)~$\mu$b.

The polarization-related systematic uncertainties were estimated in the helicity (HE) and Collins-Soper (CS) reference frames~\cite{Fac10}. The uncertainties are quoted in the frames where they are larger. Existing polarization results for $\sqrt{s}=7$ TeV at forward rapidity~\cite{Abe11}, tend to exclude a significant degree of polarization for the J/$\psi$. However, in absence of clear predictions for the $\sqrt{s}$-dependence of the effect, we prefer to quote systematic uncertainties relative to fully longitudinal ($\lambda=-1$) or transverse ($\lambda=1$) degree of polarization.
With respect to the $\sqrt{s}=7$ TeV measurement, the $\sqrt{s}=2.76$ TeV cross sections are smaller by a factor  1.59$\pm$0.50 (1.89$\pm$0.31) for the $|y|<0.9$ ($2.5<y<4$) rapidity ranges. The quoted uncertainty on the ratios is obtained by propagating the quadratic sum of statistical and systematic uncertainties (excluding the polarization-related contribution) of the two cross section values.

\begin{figure}[htbp]
\centering\resizebox{0.6\textwidth}{!}
{\includegraphics*[bb=0 0 565 546]{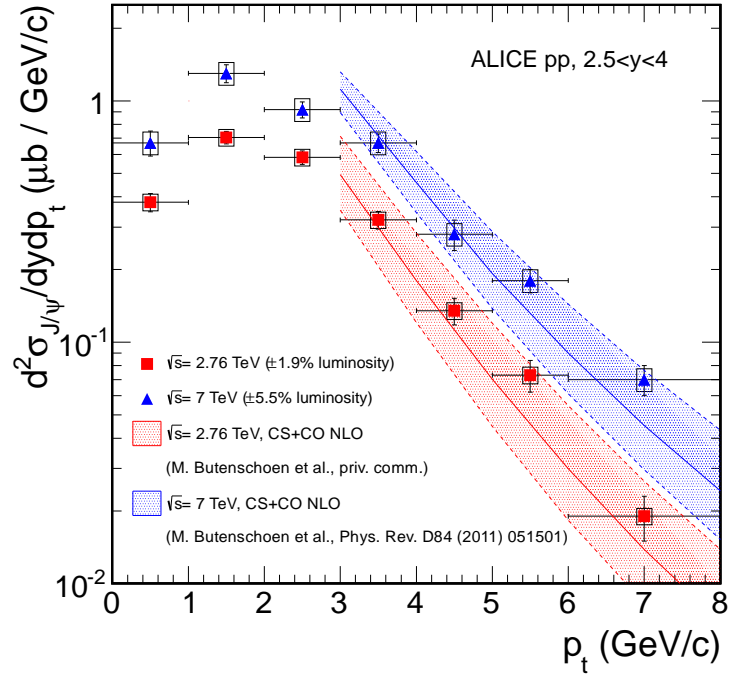}}
\caption{Double differential J/$\psi$ production cross section at $\sqrt{s}=2.76$~TeV compared to previous ALICE results at $\sqrt{s}=7$~TeV~\cite{Aam11}.
The vertical error bars represent the statistical errors
while the boxes correspond to the systematic uncertainties. The systematic uncertainties on luminosity are not included.
The results are compared with a NLO NRQCD calculation~\cite{But11a} performed in the region $p_{\rm t}>3$ GeV/$c$.}
\label{fig:4}
\end{figure}

Figure~\ref{fig:4} presents the differential cross section 
${\mathrm d}^2\sigma_{\rm J/\psi}/{\mathrm d}p_{\rm t}{\mathrm d}y$, averaged over the interval  $2.5<y<4$, for the transverse momentum range $0<p_{\rm t}<8$ GeV/$c$. The results are compared with those previously published by ALICE for $\sqrt{s}=7$ TeV, as well as, for the range $3<p_{\rm t}<8$ GeV/$c$, with the predictions of a NRQCD calculation~\cite{But11a}, which includes both colour singlet and colour octet terms at NLO. The model satisfactorily describes both sets of experimental data.

Using the results shown in Fig.~\ref{fig:4}, the mean transverse momentum for inclusive J/$\psi$ production at forward rapidity is computed by fitting ${\mathrm d}^2\sigma_{\rm J/\psi}/{\mathrm d}p_{\rm t}{\mathrm d}y$ with the function 

\begin{equation}
\frac{{\mathrm d}^2\sigma}{{\mathrm d}p_{\rm t}{\mathrm d}y}=C\frac{p_{\rm t}}{\left[1+\left(\frac{p_{\rm t}}{p_{\rm 0}}\right)^2\right]^n}
\end{equation}

with $C$, $p_{\rm 0}$ and $n$ as free parameters, as done in~\cite{Ada07}. The result, relative to the range $0<p_{\rm t}<8$ GeV/$c$, is $\langle p_{\rm t}\rangle=2.28\pm 0.07 ({\rm stat})\pm 0.04({\rm syst})$ GeV/$c$. A similar analysis carried out on the $\sqrt{s}=7$~TeV data published in~\cite{Aam11} gives $\langle p_{\rm t}\rangle=2.44\pm 0.09 ({\rm stat})\pm 0.06({\rm syst})$ GeV/$c$ for $2.5<y<4$ and $\langle p_{\rm t}\rangle=2.72\pm 0.21 ({\rm stat})\pm 0.28 ({\rm syst})$ GeV/$c$ for $|y|<0.9$ (for that data sample ${\mathrm d}^2\sigma_{\rm J/\psi}/{\mathrm d}p_{\rm t}{\mathrm d}y$ was also calculated for the dielectron analysis, in the range $0<p_{\rm t}<7$ GeV/$c$). The quoted systematic uncertainties are related to the uncorrelated systematic uncertainties for ${\mathrm d}^2\sigma/{\mathrm d}p_{\rm t}{\mathrm d}y$.

\begin{figure}[htbp]
\centering\resizebox{0.6\textwidth}{!}
{\includegraphics*[bb=0 0 565 546]{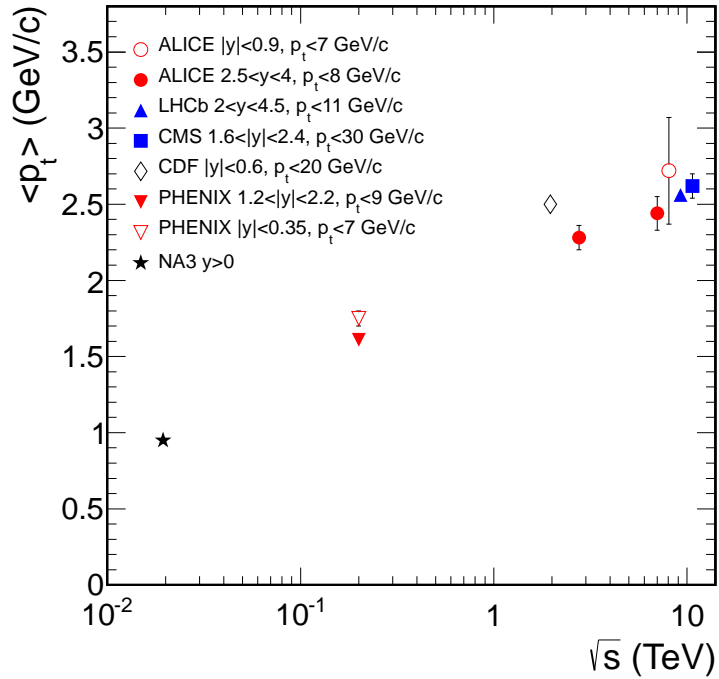}}
\caption{The $\sqrt{s}$-dependence of  $\langle p_{\rm t}\rangle$ for inclusive J/$\psi$ production, for various fixed-target and collider experiments. For the ALICE points the error bars represent the quadratic sum of statistical and systematic uncertainties. The points for $\sqrt{s}=7$ TeV have been slightly shifted to improve visibility.}
\label{fig:5}
\end{figure}

Figure~\ref{fig:5} presents the $\sqrt{s}$-dependence of the inclusive J/$\psi$ 
$\langle p_{\rm t}\rangle$, for various fixed-target and collider experiments~\cite{Aai11,Aam11,Cha11,Aco05,Ada07,Bad83}. The results show a roughly linear increase of $\langle p_{\rm t}\rangle$ with $\ln(\sqrt{s})$, with slightly larger $\langle p_{\rm t}\rangle$ values at central rapidity. The numerical values for both 
$\langle p_{\rm t}\rangle$ and $\langle p^2_{\rm t}\rangle$ are quoted in Table~\ref{tab:2}. 

\begin{table}
\caption{\label{tab:2} The 
$\langle p_{\rm t}\rangle$ and $\langle p^2_{\rm t}\rangle$ values for inclusive J/$\psi$ production measured by ALICE. Statistical and systematic uncertainties are quoted separately.}
\centering
\begin{tabular}{c|c|c}
\hline
\hline
 & $\langle p_{\rm t}\rangle$ (GeV/$c$) & $\langle p^2_{\rm T}\rangle$
 (GeV/$c$)$^2$ \\ \hline
$\sqrt{s}= 2.76$ TeV, $2.5<y<4$ & 2.28 $\pm$ 0.07 $\pm$ 0.04 & 7.06 $\pm$ 0.40 $\pm$ 0.22 \\ \hline
$\sqrt{s}= 7$ TeV, $|y|<0.9$ & 2.72 $\pm$ 0.21 $\pm$ 0.28 &  10.02 $\pm$ 1.40 $\pm$ 1.80 \\ \hline
$\sqrt{s}= 7$ TeV, $2.5<y<4$ & 2.44 $\pm$ 0.09 $\pm$ 0.06 & 8.32 $\pm$ 0.50 $\pm$ 0.35 \\ \hline
\hline
\end{tabular}
\end{table}

\begin{figure}[htbp]
\centering\resizebox{0.6\textwidth}{!}
{\includegraphics*[bb=0 0 565 546]{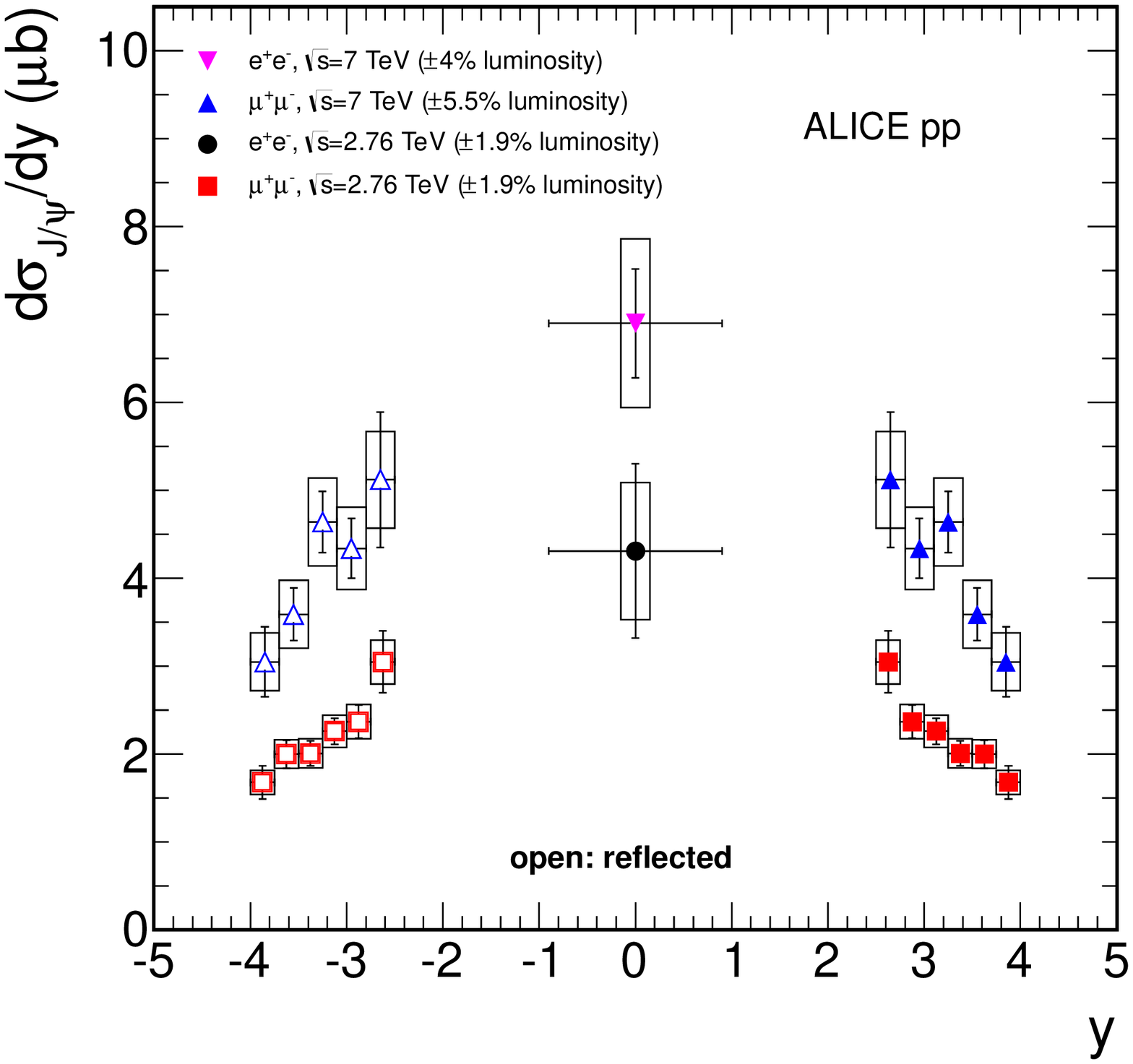}}
\caption{Differential J/$\psi$ production cross section at $\sqrt{s}=2.76$~TeV compared to previous ALICE results at $\sqrt{s}=7$~TeV~\cite{Aam11}.
The vertical error bars represent the statistical errors
while the boxes correspond to the systematic uncertainties. The systematic uncertainties on luminosity are not included.}
\label{fig:6}
\end{figure}

In Fig.~\ref{fig:6} we present the results for ${\mathrm d}\sigma_{\rm J/\psi}/{\mathrm d}y$ at $\sqrt{s}=2.76$ TeV, compared with the previously published $\sqrt{s}=7$~TeV results. The numerical values corresponding to the results presented in Fig.~\ref{fig:4} and Fig.~\ref{fig:6} are shown in Table~\ref{tab:3}, together with the number of signal events and with the values for $A\times\epsilon$.
Most sources of systematic uncertainty are common or strongly bin-to-bin correlated, except, as outlined before, the ones related to the signal extraction and to the MC inputs that are therefore quoted separately in Table~\ref{tab:3}. 

\begin{sidewaystable}
\caption{\label{tab:3}Summary of the results concerning the \jpsi\ differential cross sections for pp at $\sqrt{s}=2.76$~TeV.}
\centering
\begin{tabular}{cccccccc}
$p_{\rm t}$ & $N_{\rm J/\psi}$ & A$\times\epsilon$ & ${\mathrm d}^2\sigma_{\rm J/\psi}/{\mathrm d}p_{\rm t}{\mathrm d} y$ & \multicolumn{4}{c} {Systematic uncertainties } \\
 (GeV/$c$) &  &  & ($\mu$b/(GeV/$c$)) &  Correl. & Non-correl. & Polariz., CS & Polariz., HE \\
  &  &  &  & {($\mu$b/(GeV/$c$))} & {($\mu$b/(GeV/$c$))} & {($\mu$b/(GeV/$c$))} & {($\mu$b/(GeV/$c$))}\\
\hline
& \multicolumn{6}{c} {$2.5<y<4$} & \\  
\hline
 $[0; 1]$ & 222$\pm$19 & 0.330 & 0.380$\pm$0.033 & 0.022 & 0.021 &  $+0.074,-0.141$  &  $+0.069,-0.133$  \\
 $[1; 2]$ & 407$\pm$24 & 0.326 & 0.705$\pm$0.042 & 0.041 & 0.040 &  $+0.122,-0.271$  &  $+0.098,-0.211$  \\
 $[2; 3]$ & 343$\pm$22 & 0.332 & 0.583$\pm$0.038 & 0.034 & 0.033 &  $+0.100,-0.203$  &  $+0.069,-0.127$  \\
 $[3; 4]$ & 201$\pm$17 & 0.354 & 0.321$\pm$0.027 & 0.019 & 0.018 &  $+0.050,-0.089$  &  $+0.029,-0.047$  \\
 $[4; 5]$ & 95$\pm$12  & 0.397 & 0.135$\pm$0.017 & 0.008 & 0.008 &  $+0.014,-0.027$  &  $+0.009,-0.018$  \\
 $[5; 6]$ & 58$\pm$9   & 0.449 & 0.073$\pm$0.011 & 0.004 & 0.004 &  $+0.005,-0.011$  &  $+0.005,-0.009$  \\
 $[6; 8]$ & 34$\pm$7   & 0.507 & 0.019$\pm$0.004 & 0.001 & 0.001 &  $+0.001,-0.001$  &  $+0.001,-0.002$  \\
\hline
\hline
$y$ &  &  & ${\mathrm d}\sigma_{\rm J/\psi}/{\mathrm d} y$ ($\mu$b) & ($\mu$b) & ($\mu$b) &  ($\mu$b) &  ($\mu$b)\\
\hline
 $[-0.9; 0.9]$ & 59$\pm$14  & 0.120 & 4.31$\pm$0.99 & 0.08 & 0.77 & $+0.57,-0.81$ & $+0.65,-0.90$ \\
 $[2.5; 2.75]$ & 121$\pm$14 & 0.134 & 3.05$\pm$0.35 & 0.18 & 0.17 & $+0.67,-1.41$ & $+0.52,-1.04$ \\
 $[2.75; 3]$   & 252$\pm$20 & 0.361 & 2.37$\pm$0.19 & 0.14 & 0.13 & $+0.42,-0.84$ & $+0.39,-0.78$ \\
 $[3; 3.25]$   & 325$\pm$22 & 0.488 & 2.26$\pm$0.15 & 0.13 & 0.13 & $+0.29,-0.65$ & $+0.31,-0.61$ \\
 $[3.25; 3.5]$ & 298$\pm$21 & 0.502 & 2.01$\pm$0.14 & 0.12 & 0.11 & $+0.27,-0.54$ & $+0.21,-0.38$ \\
 $[3.5; 3.75]$ & 245$\pm$19 & 0.416 & 2.00$\pm$0.16 & 0.12 & 0.11 & $+0.33,-0.67$ & $+0.15,-0.30$ \\
 $[3.75; 4]$   & 106$\pm$12 & 0.214 & 1.68$\pm$0.19 & 0.10 & 0.09 & $+0.36,-0.69$ & $+0.16,-0.26$ \\
\end{tabular}										   
\end{sidewaystable}

\begin{figure}[htbp]
\centering\resizebox{0.6\textwidth}{!}
{\includegraphics*[bb=0 0 565 546]{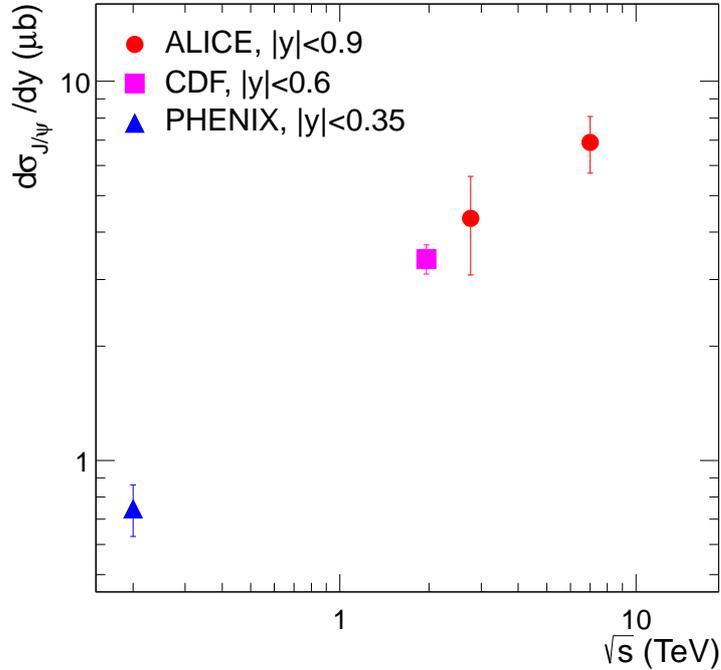}}
\caption{$\sqrt{s}$-dependence of the inclusive J/$\psi$ production cross section ${\rm d}\sigma/{\rm d}y$, at central rapidity for various collider experiments.}
\label{fig:7}
\end{figure}

The kinematic coverage of the ALICE experiment is unique among the LHC experiments due to the very good acceptance down to $p_{\rm t}=0$ at central rapidity. This feature allows a comparison of the 
$p_{\rm t}$-integrated midrapidity cross sections with those from lower energy collider experiments. The result is displayed in Fig.~\ref{fig:7}, where the  ${\mathrm d}\sigma_{\rm J/\psi}/{\mathrm d}y$ values from ALICE for the two energies are shown together with results from RHIC~\cite{Ada07} and Tevatron~\cite{Aco05} experiments, as a function of $\sqrt{s}$. 

\section{Conclusions}

The ALICE experiment has measured the inclusive J/$\psi$ production cross section for proton-proton collisions at $\sqrt{s}=2.76$ TeV, in the rapidity ranges $|y|<0.9$ and $2.5<y<4$, down to $p_{\rm t}=0$. The measured values are
$\sigma_{\rm J/\psi}(|y|<0.9)=7.75\pm 1.78({\rm{stat}})\pm 1.39({\rm{syst}})+1.16(\lambda_{\rm{HE}}=1)-1.63(\lambda_{\rm{HE}}=-1)$~$\mu$b and
$\sigma_{\rm J/\psi}(2.5<y<4)=3.34\pm 0.13({\rm{stat}})
\pm 0.27({\rm{syst}})+0.53(\lambda_{\rm{CS}}=1)-1.07(\lambda_{\rm{CS}}=-1)$~$\mu$b.
Differential cross sections in $y$ and $p_{\rm t}$ have also been measured for the forward rapidity region.
These results provide an important intermediate point between top Tevatron energy and the current maximum LHC energy. They also represent a crucial reference for the measurement of nuclear effects on J/$\psi$ production in \mbox{Pb-Pb} interactions carried out at the same 
centre-of-mass energy per nucleon-nucleon collision~\cite{Aam11J}.

%% file: acknowledgements_Jan2012.tex
The ALICE collaboration would like to thank all its engineers and technicians for their invaluable contributions to the construction of the experiment and the CERN accelerator teams for the outstanding performance of the LHC complex.
\\
The ALICE collaboration acknowledges the following funding agencies for their support in building and
running the ALICE detector:
 \\
Calouste Gulbenkian Foundation from Lisbon and Swiss Fonds Kidagan, Armenia;
 \\
Conselho Nacional de Desenvolvimento Cient\'{\i}fico e Tecnol\'{o}gico (CNPq), Financiadora de Estudos e Projetos (FINEP),
Funda\c{c}\~{a}o de Amparo \`{a} Pesquisa do Estado de S\~{a}o Paulo (FAPESP);
 \\
National Natural Science Foundation of China (NSFC), the Chinese Ministry of Education (CMOE)
and the Ministry of Science and Technology of China (MSTC);
 \\
Ministry of Education and Youth of the Czech Republic;
 \\
Danish Natural Science Research Council, the Carlsberg Foundation and the Danish National Research Foundation;
 \\
The European Research Council under the European Community's Seventh Framework Programme;
 \\
Helsinki Institute of Physics and the Academy of Finland;
 \\
French CNRS-IN2P3, the `Region Pays de Loire', `Region Alsace', `Region Auvergne' and CEA, France;
 \\
German BMBF and the Helmholtz Association;
\\
General Secretariat for Research and Technology, Ministry of
Development, Greece;
\\
Hungarian OTKA and National Office for Research and Technology (NKTH);
 \\
Department of Atomic Energy and Department of Science and Technology of the Government of India;
 \\
Istituto Nazionale di Fisica Nucleare (INFN) of Italy;
 \\
MEXT Grant-in-Aid for Specially Promoted Research, Ja\-pan;
 \\
Joint Institute for Nuclear Research, Dubna;
 \\
National Research Foundation of Korea (NRF);
 \\
CONACYT, DGAPA, M\'{e}xico, ALFA-EC and the HELEN Program (High-Energy physics Latin-American--European Network);
 \\
Stichting voor Fundamenteel Onderzoek der Materie (FOM) and the Nederlandse Organisatie voor Wetenschappelijk Onderzoek (NWO), Netherlands;
 \\
Research Council of Norway (NFR);
 \\
Polish Ministry of Science and Higher Education;
 \\
National Authority for Scientific Research - NASR (Autoritatea Na\c{t}ional\u{a} pentru Cercetare \c{S}tiin\c{t}ific\u{a} - ANCS);
 \\
Federal Agency of Science of the Ministry of Education and Science of Russian Federation, International Science and
Technology Center, Russian Academy of Sciences, Russian Federal Agency of Atomic Energy, Russian Federal Agency for Science and Innovations and CERN-INTAS;
 \\
Ministry of Education of Slovakia;
 \\
Department of Science and Technology, South Africa;
 \\
CIEMAT, EELA, Ministerio de Educaci\'{o}n y Ciencia of Spain, Xunta de Galicia (Conseller\'{\i}a de Educaci\'{o}n),
CEA\-DEN, Cubaenerg\'{\i}a, Cuba, and IAEA (International Atomic Energy Agency);
 \\
Swedish Research Council (VR) and Knut $\&$ Alice Wallenberg
Foundation (KAW);
 \\
Ukraine Ministry of Education and Science;
 \\
United Kingdom Science and Technology Facilities Council (STFC);
 \\
The United States Department of Energy, the United States National
Science Foundation, the State of Texas, and the State of Ohio.

%% file: authorlist_17Feb2012.tex
\begingroup
\small
\begin{flushleft}
B.~Abelev\Irefn{org1234}\And
J.~Adam\Irefn{org1274}\And
D.~Adamov\'{a}\Irefn{org1283}\And
A.M.~Adare\Irefn{org1260}\And
M.M.~Aggarwal\Irefn{org1157}\And
G.~Aglieri~Rinella\Irefn{org1192}\And
A.G.~Agocs\Irefn{org1143}\And
A.~Agostinelli\Irefn{org1132}\And
S.~Aguilar~Salazar\Irefn{org1247}\And
Z.~Ahammed\Irefn{org1225}
% ...
\renewcommand\labelenumi{\textsuperscript{\theenumi}~}
\section*{Affiliation notes}
\renewcommand\theenumi{\roman{enumi}}
\begin{Authlist}
\item \Adef{M.V.Lomonosov Moscow State University, D.V.Skobeltsyn Institute of Nuclear Physics, Moscow, Russia}Also at: M.V.Lomonosov Moscow State University, D.V.Skobeltsyn Institute of Nuclear Physics, Moscow, Russia
\item \Adef{Institute of Nuclear Sciences, Belgrade, Serbia}Also at: "Vin\v{c}a" Institute of Nuclear Sciences, Belgrade, Serbia
\end{Authlist}
\section*{Collaboration Institutes}
\renewcommand\theenumi{\arabic{enumi}~}
\begin{Authlist}
\item \Idef{org1279}Benem\'{e}rita Universidad Aut\'{o}noma de Puebla, Puebla, Mexico
\item \Idef{org1220}Bogolyubov Institute for Theoretical Physics, Kiev, Ukraine
\item \Idef{org1262}Budker Institute for Nuclear Physics, Novosibirsk, Russia
\item \Idef{org1292}California Polytechnic State University, San Luis Obispo, California, United States
\item \Idef{org14939}Centre de Calcul de l'IN2P3, Villeurbanne, France
\item \Idef{org1197}Centro de Aplicaciones Tecnol\'{o}gicas y Desarrollo Nuclear (CEADEN), Havana, Cuba
\item \Idef{org1242}Centro de Investigaciones Energ\'{e}ticas Medioambientales y Tecnol\'{o}gicas (CIEMAT), Madrid, Spain
\item \Idef{org1244}Centro de Investigaci\'{o}n y de Estudios Avanzados (CINVESTAV), Mexico City and M\'{e}rida, Mexico
\item \Idef{org1335}Centro Fermi -- Centro Studi e Ricerche e Museo Storico della Fisica ``Enrico Fermi'', Rome, Italy
\item \Idef{org17347}Chicago State University, Chicago, United States
\item \Idef{org1288}Commissariat \`{a} l'Energie Atomique, IRFU, Saclay, France
\item \Idef{org1294}Departamento de F\'{\i}sica de Part\'{\i}culas and IGFAE, Universidad de Santiago de Compostela, Santiago de Compostela, Spain
\item \Idef{org1106}Department of Physics Aligarh Muslim University, Aligarh, India
\item \Idef{org1121}Department of Physics and Technology, University of Bergen, Bergen, Norway
\item \Idef{org1162}Department of Physics, Ohio State University, Columbus, Ohio, United States
\item \Idef{org1300}Department of Physics, Sejong University, Seoul, South Korea
\item \Idef{org1268}Department of Physics, University of Oslo, Oslo, Norway
\item \Idef{org1145}Dipartimento di Fisica dell'Universit\`{a} and Sezione INFN, Cagliari, Italy
\item \Idef{org1270}Dipartimento di Fisica dell'Universit\`{a} and Sezione INFN, Padova, Italy
\item \Idef{org1315}Dipartimento di Fisica dell'Universit\`{a} and Sezione INFN, Trieste, Italy
\item \Idef{org1132}Dipartimento di Fisica dell'Universit\`{a} and Sezione INFN, Bologna, Italy
\item \Idef{org1285}Dipartimento di Fisica dell'Universit\`{a} `La Sapienza' and Sezione INFN, Rome, Italy
\item \Idef{org1154}Dipartimento di Fisica e Astronomia dell'Universit\`{a} and Sezione INFN, Catania, Italy
\item \Idef{org1290}Dipartimento di Fisica `E.R.~Caianiello' dell'Universit\`{a} and Gruppo Collegato INFN, Salerno, Italy
\item \Idef{org1312}Dipartimento di Fisica Sperimentale dell'Universit\`{a} and Sezione INFN, Turin, Italy
\item \Idef{org1103}Dipartimento di Scienze e Tecnologie Avanzate dell'Universit\`{a} del Piemonte Orientale and Gruppo Collegato INFN, Alessandria, Italy
\item \Idef{org1114}Dipartimento Interateneo di Fisica `M.~Merlin' and Sezione INFN, Bari, Italy
\item \Idef{org1237}Division of Experimental High Energy Physics, University of Lund, Lund, Sweden
\item \Idef{org1192}European Organization for Nuclear Research (CERN), Geneva, Switzerland
\item \Idef{org1227}Fachhochschule K\"{o}ln, K\"{o}ln, Germany
\item \Idef{org1122}Faculty of Engineering, Bergen University College, Bergen, Norway
\item \Idef{org1136}Faculty of Mathematics, Physics and Informatics, Comenius University, Bratislava, Slovakia
\item \Idef{org1274}Faculty of Nuclear Sciences and Physical Engineering, Czech Technical University in Prague, Prague, Czech Republic
\item \Idef{org1229}Faculty of Science, P.J.~\v{S}af\'{a}rik University, Ko\v{s}ice, Slovakia
\item \Idef{org1184}Frankfurt Institute for Advanced Studies, Johann Wolfgang Goethe-Universit\"{a}t Frankfurt, Frankfurt, Germany
\item \Idef{org1215}Gangneung-Wonju National University, Gangneung, South Korea
\item \Idef{org1212}Helsinki Institute of Physics (HIP) and University of Jyv\"{a}skyl\"{a}, Jyv\"{a}skyl\"{a}, Finland
\item \Idef{org1203}Hiroshima University, Hiroshima, Japan
\item \Idef{org1329}Hua-Zhong Normal University, Wuhan, China
\item \Idef{org1254}Indian Institute of Technology, Mumbai, India
\item \Idef{org36378}Indian Institute of Technology Indore (IIT), Indore, India
\item \Idef{org1266}Institut de Physique Nucl\'{e}aire d'Orsay (IPNO), Universit\'{e} Paris-Sud, CNRS-IN2P3, Orsay, France
\item \Idef{org1277}Institute for High Energy Physics, Protvino, Russia
\item \Idef{org1249}Institute for Nuclear Research, Academy of Sciences, Moscow, Russia
\item \Idef{org1320}Nikhef, National Institute for Subatomic Physics and Institute for Subatomic Physics of Utrecht University, Utrecht, Netherlands
\item \Idef{org1250}Institute for Theoretical and Experimental Physics, Moscow, Russia
\item \Idef{org1230}Institute of Experimental Physics, Slovak Academy of Sciences, Ko\v{s}ice, Slovakia
\item \Idef{org1127}Institute of Physics, Bhubaneswar, India
\item \Idef{org1275}Institute of Physics, Academy of Sciences of the Czech Republic, Prague, Czech Republic
\item \Idef{org1139}Institute of Space Sciences (ISS), Bucharest, Romania
\item \Idef{org27399}Institut f\"{u}r Informatik, Johann Wolfgang Goethe-Universit\"{a}t Frankfurt, Frankfurt, Germany
\item \Idef{org1185}Institut f\"{u}r Kernphysik, Johann Wolfgang Goethe-Universit\"{a}t Frankfurt, Frankfurt, Germany
\item \Idef{org1177}Institut f\"{u}r Kernphysik, Technische Universit\"{a}t Darmstadt, Darmstadt, Germany
\item \Idef{org1256}Institut f\"{u}r Kernphysik, Westf\"{a}lische Wilhelms-Universit\"{a}t M\"{u}nster, M\"{u}nster, Germany
\item \Idef{org1246}Instituto de Ciencias Nucleares, Universidad Nacional Aut\'{o}noma de M\'{e}xico, Mexico City, Mexico
\item \Idef{org1247}Instituto de F\'{\i}sica, Universidad Nacional Aut\'{o}noma de M\'{e}xico, Mexico City, Mexico
\item \Idef{org23333}Institut of Theoretical Physics, University of Wroclaw
\item \Idef{org1308}Institut Pluridisciplinaire Hubert Curien (IPHC), Universit\'{e} de Strasbourg, CNRS-IN2P3, Strasbourg, France
\item \Idef{org1182}Joint Institute for Nuclear Research (JINR), Dubna, Russia
\item \Idef{org1143}KFKI Research Institute for Particle and Nuclear Physics, Hungarian Academy of Sciences, Budapest, Hungary
\item \Idef{org1199}Kirchhoff-Institut f\"{u}r Physik, Ruprecht-Karls-Universit\"{a}t Heidelberg, Heidelberg, Germany
\item \Idef{org20954}Korea Institute of Science and Technology Information, Daejeon, South Korea
\item \Idef{org1160}Laboratoire de Physique Corpusculaire (LPC), Clermont Universit\'{e}, Universit\'{e} Blaise Pascal, CNRS--IN2P3, Clermont-Ferrand, France
\item \Idef{org1194}Laboratoire de Physique Subatomique et de Cosmologie (LPSC), Universit\'{e} Joseph Fourier, CNRS-IN2P3, Institut Polytechnique de Grenoble, Grenoble, France
\item \Idef{org1187}Laboratori Nazionali di Frascati, INFN, Frascati, Italy
\item \Idef{org1232}Laboratori Nazionali di Legnaro, INFN, Legnaro, Italy
\item \Idef{org1125}Lawrence Berkeley National Laboratory, Berkeley, California, United States
\item \Idef{org1234}Lawrence Livermore National Laboratory, Livermore, California, United States
\item \Idef{org1251}Moscow Engineering Physics Institute, Moscow, Russia
\item \Idef{org1140}National Institute for Physics and Nuclear Engineering, Bucharest, Romania
\item \Idef{org1165}Niels Bohr Institute, University of Copenhagen, Copenhagen, Denmark
\item \Idef{org1109}Nikhef, National Institute for Subatomic Physics, Amsterdam, Netherlands
\item \Idef{org1283}Nuclear Physics Institute, Academy of Sciences of the Czech Republic, \v{R}e\v{z} u Prahy, Czech Republic
\item \Idef{org1264}Oak Ridge National Laboratory, Oak Ridge, Tennessee, United States
\item \Idef{org1189}Petersburg Nuclear Physics Institute, Gatchina, Russia
\item \Idef{org1170}Physics Department, Creighton University, Omaha, Nebraska, United States
\item \Idef{org1157}Physics Department, Panjab University, Chandigarh, India
\item \Idef{org1112}Physics Department, University of Athens, Athens, Greece
\item \Idef{org1152}Physics Department, University of Cape Town, iThemba LABS, Cape Town, South Africa
\item \Idef{org1209}Physics Department, University of Jammu, Jammu, India
\item \Idef{org1207}Physics Department, University of Rajasthan, Jaipur, India
\item \Idef{org1200}Physikalisches Institut, Ruprecht-Karls-Universit\"{a}t Heidelberg, Heidelberg, Germany
\item \Idef{org1325}Purdue University, West Lafayette, Indiana, United States
\item \Idef{org1281}Pusan National University, Pusan, South Korea
\item \Idef{org1176}Research Division and ExtreMe Matter Institute EMMI, GSI Helmholtzzentrum f\"ur Schwerionenforschung, Darmstadt, Germany
\item \Idef{org1334}Rudjer Bo\v{s}kovi\'{c} Institute, Zagreb, Croatia
\item \Idef{org1298}Russian Federal Nuclear Center (VNIIEF), Sarov, Russia
\item \Idef{org1252}Russian Research Centre Kurchatov Institute, Moscow, Russia
\item \Idef{org1224}Saha Institute of Nuclear Physics, Kolkata, India
\item \Idef{org1130}School of Physics and Astronomy, University of Birmingham, Birmingham, United Kingdom
\item \Idef{org1338}Secci\'{o}n F\'{\i}sica, Departamento de Ciencias, Pontificia Universidad Cat\'{o}lica del Per\'{u}, Lima, Peru
\item \Idef{org1316}Sezione INFN, Trieste, Italy
\item \Idef{org1271}Sezione INFN, Padova, Italy
\item \Idef{org1313}Sezione INFN, Turin, Italy
\item \Idef{org1286}Sezione INFN, Rome, Italy
\item \Idef{org1146}Sezione INFN, Cagliari, Italy
\item \Idef{org1133}Sezione INFN, Bologna, Italy
\item \Idef{org1115}Sezione INFN, Bari, Italy
\item \Idef{org1155}Sezione INFN, Catania, Italy
\item \Idef{org1322}Soltan Institute for Nuclear Studies, Warsaw, Poland
\item \Idef{org36377}Nuclear Physics Group, STFC Daresbury Laboratory, Daresbury, United Kingdom
\item \Idef{org1258}SUBATECH, Ecole des Mines de Nantes, Universit\'{e} de Nantes, CNRS-IN2P3, Nantes, France
\item \Idef{org1304}Technical University of Split FESB, Split, Croatia
\item \Idef{org1168}The Henryk Niewodniczanski Institute of Nuclear Physics, Polish Academy of Sciences, Cracow, Poland
\item \Idef{org17361}The University of Texas at Austin, Physics Department, Austin, TX, United States
\item \Idef{org1173}Universidad Aut\'{o}noma de Sinaloa, Culiac\'{a}n, Mexico
\item \Idef{org1296}Universidade de S\~{a}o Paulo (USP), S\~{a}o Paulo, Brazil
\item \Idef{org1149}Universidade Estadual de Campinas (UNICAMP), Campinas, Brazil
\item \Idef{org1239}Universit\'{e} de Lyon, Universit\'{e} Lyon 1, CNRS/IN2P3, IPN-Lyon, Villeurbanne, France
\item \Idef{org1205}University of Houston, Houston, Texas, United States
\item \Idef{org20371}University of Technology and Austrian Academy of Sciences, Vienna, Austria
\item \Idef{org1222}University of Tennessee, Knoxville, Tennessee, United States
\item \Idef{org1310}University of Tokyo, Tokyo, Japan
\item \Idef{org1318}University of Tsukuba, Tsukuba, Japan
\item \Idef{org21360}Eberhard Karls Universit\"{a}t T\"{u}bingen, T\"{u}bingen, Germany
\item \Idef{org1225}Variable Energy Cyclotron Centre, Kolkata, India
\item \Idef{org1306}V.~Fock Institute for Physics, St. Petersburg State University, St. Petersburg, Russia
\item \Idef{org1323}Warsaw University of Technology, Warsaw, Poland
\item \Idef{org1179}Wayne State University, Detroit, Michigan, United States
\item \Idef{org1260}Yale University, New Haven, Connecticut, United States
\item \Idef{org1332}Yerevan Physics Institute, Yerevan, Armenia
\item \Idef{org15649}Yildiz Technical University, Istanbul, Turkey
\item \Idef{org1301}Yonsei University, Seoul, South Korea
\item \Idef{org1327}Zentrum f\"{u}r Technologietransfer und Telekommunikation (ZTT), Fachhochschule Worms, Worms, Germany
\end{Authlist}
\endgroup